\documentclass[12pt]{iopart}
\usepackage{iopams} 
\usepackage{epsfig} 
\begin{document}

\title[Nonlinear response of a driven vibrating nanobeam in the quantum regime]
{Nonlinear response of a driven vibrating nanobeam in the
quantum regime}
 
\author{ V Peano and  M Thorwart}

\address{Institut f\"ur Theoretische Physik, 
Heinrich-Heine-Universit\"at D\"usseldorf, D-40225 D\"usseldorf, Germany}

\submitto{\NJP}

\begin{abstract}
We analytically investigate the nonlinear response of a damped doubly clamped
nanomechanical beam under static longitudinal compression 
which is excited to transverse vibrations.  
Starting from a continuous elasticity model for the beam, we consider
the dynamics of the beam close to the Euler buckling instability. 
There, the fundamental transverse mode dominates and 
a quantum mechanical time-dependent 
effective single particle Hamiltonian for its amplitude 
can be derived. In addition, we include the influence of a
dissipative Ohmic or super-Ohmic environment. In the rotating frame, a 
Markovian master equation is derived which includes also the effect of
the time-dependent driving in a non-trivial way.  
The quasienergies of the pure system 
show  multiple avoided level crossings corresponding to multiphonon
transitions in the resonator. Around the resonances, the 
master equation is solved analytically using 
Van Vleck perturbation theory. 
Their lineshapes are calculated resulting in simple expressions. 
We find the general solution for the 
multiple multiphonon resonances and, most interestingly, 
a bath-induced transition from a resonant to an antiresonant behavior
of the nonlinear response. 
\end{abstract}
%
% 05.45.-a 	Nonlinear dynamics and nonlinear dynamical systems (
% 03.65.Yz 	Decoherence; open systems; quantum statistical methods 
%              (see also 03.67.Pp in quantum information; 
%              for decoherence in Bose-Einstein condensates, see 03.75.Gg)
% 62.25.+g 	Mechanical properties of nanoscale materials
% 62.30.+d 	Mechanical and elastic waves; vibrations (see also 43.40.+s 
%               Structural acoustics and vibration; 46.40.-f Vibrations and 
%               mechanical waves in continuum mechanics of solids)
% 62.40.+i 	Anelasticity, internal friction, stress relaxation, and 
%               mechanical resonances (see also 81.40.Jj Elasticity and anelasticity) 
%
% 63.22.+m 	Phonons or vibrational states in low-dimensional structures 
%               and nanoscale materials
%%%%%%%%%%%%%%%
\maketitle
\section{Introduction}
The experimental realization of nanoscale resonators which show
quantum mechanical behavior 
\cite{Craighead00,Roukes01,Cleland03,Blencowe04,Blencowe05} 
is currently on the schedule of several
research groups worldwide and poses a rather non-trivial task. 
Important key experiments on the way to
this goal have already been reported in the literature 
\cite{Roukes96,Cleland98,Nguyen99,Kroemmer00,
Erbe00,Beil00,Cleland01,Buks01,Ming03,
Knobel03,Husain03,LaHaye04,Mohanty05,Aldridge05,Ono05} and are also reviewed in this
Focus Issue. Most techniques to reveal the quantum behavior 
so far address the linear response in form of the  amplitude  
of the transverse vibrations 
 of the nanobeam around its
eigenfrequency. The goal is to  excite only a few energy quanta in a
 resonator held at low temperature. To measure the response,  
the ultimate goal of the experiments is to
increase the resolution of the position  measurement 
to the quantum limit \cite{Beil00,LaHaye04,Mohanty05,
Schwab05,Gaidarzhy05}. As the response of a damped linear quantum oscillator
has the same simple Lorentzian shape as the one of a damped linear classical 
oscillator \cite{Weiss99}, a unique identification of the
``quantumness'' of a nanoresonator in the linear regime 
can sometimes be difficult. 

One possible alternative is to study the nonlinear response of the
nanoresonator which has been excited to its nonlinear regime. 
A macroscopic beam which is clamped at its ends and which is strongly 
excited to transverse vibrations displays the properties of the Duffing oscillator being 
a simple  damped driven oscillator with a (cubic) nonlinear restoring force 
\cite{Nayfeh}. Its nonlinear response displays rich physical
properties including a driving induced bistability, hysteresis,
harmonic mixing and
chaos \cite{Nayfeh,Holmes,Jackson}. 
The nonlinear response of (still classical) nanoscale resonators has
been measured in recent experiments
\cite{Kroemmer00,Erbe00,Beil00,Husain03}. In the range of  weak
excitations, the standard linear response arises while for increasing
driving, the characteristic response
curve of a classical Duffing oscillator has been identified. 

No signatures of a quantum behavior in the nonlinear response of
 realized nanobeams have been
reported up to present. One reason is that a  nanomechanical resonator is exposed to a
 variety of intrinsic as well as extrinsic 
damping mechanisms depending on the details 
of the fabrication procedure, the experimental conditions and the used 
materials \cite{Ono05,Mohanty02,Hutchinson04,Liu05}. Possible extrinsic mechanisms include 
clamping losses due to the strain at the connections to the support structure, heating, 
coupling to higher vibrational modes, friction due to the surrounding gas, 
nonlinear effects, 
thermoelastic losses due to propagating acoustic waves, 
surface roughness, extrinsic noise sources, 
dislocations, and other material-dependent properties. 
An important internal mechanism is the interaction with localized
crystal defects.
Controlling this variety of damping sources is one of the major tasks
to be solved to reveal quantum mechanical features. Recent measurement show
that in the so far realized devices based on silicon and diamond
structures, damping has been rather strong at low frequencies 
  \cite{Mohanty02,Hutchinson04} indicating even sub-Ohmic-type 
 damping \cite{Weiss99} which would make it difficult to 
 observe quantum effects at all. However, using freely suspended carbon
 nanotubes \cite{Sazonova04,Sapmaz05}  
 instead could reduce damping at low frequencies due to the
 more regular structure of the long molecules which can be produced in
 a very clean manner. Further experimental work is required to clarify
 this point and to optimize the experimental conditions. 

Nanoscale nonlinear resonators in the quantum regime have been investigated
theoretically starting from microscopic models based on 
elasticity theory for the beam \cite{Carr01,Werner04,Peano04}. Carr,
Lawrence and Wybourne have considered an elastic bar under
static longitudinal compression beyond the Euler instability 
leading to two stable equilibrium positions
around which the transverse vibrations of
the beam occur. It turned out that quantum tunneling between the two
minima is in principle
possible in silicon beams and carbon nanotubes. However, the strain
has to be controlled with extreme  accuracy and  the quantum
fluctuations in position are of the order of $0.1$ \AA. 
The detection of such small lengths certainly is challenging. However, 
a possible method to increase the resolution could be
the use of the phenomenon of 
stochastic resonance \cite{HanggiSR} for a coherent signal amplification 
 of the nonlinear response of nanomechanical resonators in their
 bistable regime \cite{Badzey05}. 
Werner and Zwerger \cite{Werner04} have studied a similar setup close
to the Euler buckling instability which occurs at a critical strain 
$\epsilon_c$. There, the frequency of the
fundamental mode vanishes and quartic terms in the Lagrangian have to
be taken into account. An effective Hamiltonian has been derived for
the amplitude of the fundamental mode being the dynamical variable
which moves in an anharmonic potential. Depending on the strain
$\epsilon$ being below ($\epsilon < \epsilon_c$) or above 
($\epsilon > \epsilon_c$) the
critical value, a monostable or bistable situation can be created. 
The conditions for macroscopic quantum tunneling  to occur have been
estimated for the bistable case. 
In order to measure single-phonon transitions in a nanoresonator, it has 
been proposed to use its anharmonicity together with a second nanoresonator acting as 
a transducer for the phonon number in the first one \cite{Santamore04}. In this way, 
the measured signal being the induced current is directly proportional 
to the position of the read-out oscillator.   

In Ref.\ \cite{Peano04}, we have considered a similar setup but
restricted to the  {\em statically monostable\/}  case below the Euler
instability, i.e., for $\epsilon < \epsilon_c$. In addition, we
have allowed for a time-dependent periodic driving force $F(t)$ such that an
effective monostable quantum Duffing oscillator arises. Possible
origin of the driving can be the magneto-motive force when an ac
current is applied and the beam is placed in a transverse magnetic
field. Moreover, a 
(weak) influence of the environment has been modeled phenomenologically by
a simple Ohmic harmonic bath. The nonlinear response has been determined 
{\em numerically} from 
solving a Born-Markovian master equation for the reduced 
density operator of  the system after the  bath has been traced out. 
We have identified discrete multiphonon transitions as well as
macroscopic quantum tunneling of the fundamental mode amplitude
between the two stable states in the {\em driving induced\/} bistability. 
Moreover, a peculiar multiphonon antiresonant behavior has been 
found in the numerical results for the damped system \cite{Peano05}. 
The discrete multiphonon \mbox{(anti-)resonances} are a typical signature 
of quantum mechanical behavior \cite{Peano04,Peano05} and are absent in 
the corresponding classical model of the standard Duffing oscillator 
\cite{Nayfeh,Holmes,Jackson}, also when thermal fluctuations are 
included \cite{Datta01}. 

While we have approached the problem in Refs.\ \cite{Peano04,Peano05}  
by  numerical means, we present in this work a complete {\em analytical} 
investigation of the dynamics of the quantum
Duffing oscillator. We intend to elucidate the mechanism behind the 
reported \cite{Peano05} bath-induced transition from the resonant to the antiresonant 
nonlinear response of the nanobeam. This is achieved by solving 
a Born-Markovian master 
equation for the reduced density operator in the rotating frame. 
Within the rotating wave approximation (RWA), a simplified system Hamiltonian
follows whose eigenstates are the quasienergy states. The
corresponding quasienergies show avoided level crossings when the
driving frequency is varied. They correspond to multiphonon
transitions occurring in the resonator. Moreover, we include the
dissipative influence of an environment and find that the dynamics
around the avoided quasienergy level crossings is well described by a
simplified master equation involving only a few quasienergy states. 
Around the anticrossings, we find resonant as well as antiresonant
nonlinear responses depending on the damping strength.  The underlying
mechanism is worked out in the perturbative regime of weak
nonlinearity, weak driving and weak damping. There, Van Vleck
perturbation theory allows to obtain the quasienergies and the
quasienergy states analytically. The master equation can then be
solved in the stationary limit and subsequently, the line shapes of
the resonant as well as the antiresonant nonlinear
response can be calculated. 

 The problem of a driven quantum oscillator with a quartic
 nonlinearity  has been 
 investigated theoretically in earlier works in various contexts. 
 In the context of the
 radiative excitation of polyatomic molecules \cite{Larsen76}, 
 Larsen and Bloembergen have calculated the wave-functions for the
 coherent multiphoton Rabi precession between two discrete levels for a
 collisionless model. More recently, also Dykman and Fistul \cite{Dykman05} have
 considered the bare nonlinear Hamiltonian under the rotating wave
 approximation. 
 Drummond and Walls \cite{Drummond80} have
 investigated a similar system occurring for the case of a coherently
 driven dispersive cavity including a cubic nonlinearity. Photon
 bunching and antibunching have been predicted upon solving the
 corresponding Fokker-Planck equation. Vogel and Risken 
 \cite{Vogel88} have calculated the tunneling rates for the
 Drummond-Walls model by use of continued fraction methods. 
 Dmitriev, D'yakonov and Ioffe \cite{Dyakonov86} 
 have calculated the tunneling and
 thermal transition rates for the case when the associated times are
 large.  
  Dykman and Smelyanskii \cite{Dykman88} have calculated the
 probability of transitions between the stable states in a
 quasi-classical approximation in the thermally activated regime.  
 Recently, the role of the detector (in this case, a photon detector)
 has been studied for the quantum Duffing oscillator in the chaotic 
 regime \cite{Everitt05}. The power spectra of the detected photons
 carry information on the underlying dynamics of the nonlinear
 oscillator and can be used to distinguish its different modes.  
 However, the line-shape of the multi-phonon resonance which is the 
 central object for studying the nonlinear response remained
 unaddressed so far. In addition, we start from a microscopic
 Hamiltonian for the bath and present  a fully analytical 
 treatment of system and environment in the deep quantum 
 regime of weak coupling.  

Our paper is structured such that we introduce the elasticity model
for the doubly clamped nanobeam, derive the effective quartic 
Hamiltonian and discuss the model for damping 
in Section \ref{modelsystem}. Then, we discuss the coherent dynamics
of the pure system in terms of the RWA 
and the Van Vleck perturbation method in Section 
\ref{coherent}. The dissipative dynamics is studied in Section 
\ref{dissipative}, while the observables are defined in Section 
\ref{observables}. The solution for the line shapes are given in
Section \ref{sol} before  the final conclusions are drawn in Section 
\ref{concl}.  

\section{Model for the driven nanoresonator\label{modelsystem}}
We consider a freely suspended nanomechanical beam of total length $L$ and 
mass density $\sigma=m/L$ which 
is clamped at both ends ({\em doubly clamped boundary conditions}) and which is 
characterized by its bending rigidity $\mu=Y I$ being the product 
of Young's elasticity modulus $Y$ and the moment of inertia $I$.  
In addition, we allow for a mechanical force $F_0>0$ which 
compresses the beam in longitudinal direction. Moreover, 
the beam is excited to transverse vibrations by a time-dependent 
driving  field $F(t) = \tilde{f} \cos (\omega_{\rm ex} t)$. 
In a classical description, the transverse deflection $\phi(s,t)$ 
characterizes the beam completely, where $0\le s \le L$. Then, the Lagrangian of the 
vibrating beam follows from elasticity theory as \cite{Werner04}
\begin{equation}
{\cal L} (\phi, \dot{\phi}, t)= 
\int_0^L ds \left[ \frac{\sigma}{2} \dot{\phi}^2 - \frac{\mu}{2} 
\frac{\phi^{\prime\prime 2}}{1 - \phi^{\prime 2}} -F_0 
\left( \sqrt{1-\phi^{\prime 2}}-1\right) + F(t) \phi 
\right] \, .
\end{equation}
Before we study the dynamics of the driven beam, we consider first the undriven 
system with $F(t) \equiv 0$. 

For the case of small deflections $|\phi^\prime (s)|
 \ll 1$, the Lagrangian can be linearized and the 
time-dependent Euler-Lagrange equations 
can be solved by the eigenfunctions $\phi(s,t) = \sum_n \phi_n(s,t) = 
\sum_n {\cal A}_n(t) g_n(s)$, where $g_n(s)$ are the normal modes which follow 
as the solution of the characteristic equation. For the doubly clamped nano-beam, we 
have $\phi(0)=\phi(L)=0$ and $\phi^\prime(0)=\phi^\prime(L)=0$. However, 
it turns out that this situation is closely related to the simpler case 
that the nano-beam is also fixed at both ends but its ends can move such that the 
bending moments at the ends vanish, i.e., 
$\phi(0)=\phi(L)=0$ and $\phi^{\prime\prime} (0)=\phi^{\prime \prime} (L)=0$ 
({\it free boundary conditions}). For the case of 
free boundary conditions, the characteristic equation yields the 
normal modes $g_n^{\rm free}(s) = \sin (n\pi s / L)$ and the 
corresponding frequency of the $n-$th mode  follows as 
\begin{equation}
\omega_n^{\rm free} = \left( \frac{\mu (n \pi / L)^2 - F_0}{\sigma} \right)^{1/2} \frac{n\pi}{L} \, .
\end{equation}
At the critical force $F_c = \mu (\pi / L)^2$, the fundamental frequency 
$\omega_1^{\rm free} (F_0 \rightarrow F_c)$ 
vanishes as $\sqrt{\epsilon}$, where $\epsilon = (F_c-F_0)/ F_c$ is the distance 
to the critical force, and the well-known Euler instability occurs. 
%In this regime, a frequency scale 
%$\omega_0$ can be defined with $\omega_1(F_0 \rightarrow F_c) = \sqrt{\epsilon} \omega_0$, 
%with $\omega_0 = \sqrt{\mu/\sigma} (\pi/L)^2$, where $. 
%

For the case of doubly clamped boundary conditions, the characteristic equation yields
 a transcendental equation for the normal modes which cannot be solved analytically. 
However, close to the Euler instability $F_0 \rightarrow F_c$, the situation simplifies 
again. After expanding, one finds for the fundamental frequency 
$\omega_1(F_0 \rightarrow F_c) = \sqrt{\epsilon} \omega_0$, with the 
frequency scale $\omega_0= (4/\sqrt{3}) \sqrt{\mu/\sigma} (\pi/L)^2$. 
Approaching the Euler instability, the frequencies of 
the higher modes $\omega_{n\ge 2}$ remain finite, while the fundamental frequency 
$\omega_1(F_0 \rightarrow F_c)$ vanishes again like $\sqrt{\epsilon}$. Hence, the dynamics at low energies 
close to the Euler instability will be dominated by the fundamental mode alone which 
simplifies the treatment of the nonlinear case, see below. The 
fundamental mode $g_1(s)$ can also be expanded close to the Euler instability and one 
obtains in zero-th order in $\epsilon$ 
\begin{equation}
g_1(s) \simeq \sin^2 \left( \frac{\pi s}{L} \right) \, . 
\end{equation}
\subsection{Effective single-particle Hamiltonian\label{effhamil}}

Since the fundamental mode vanishes when $F_0 \rightarrow F_c$, one has to include the 
contributions beyond the quadratic terms $\propto \phi^{\prime 2}, \phi^{\prime \prime 2}$ of the 
transverse deflections in the Lagrangian. The next higher order is quartic and 
yields terms $\propto \phi^{\prime 4}, \phi^{\prime 2} \phi^{\prime \prime 2}$. Inserting 
again the normal mode expansion in the Lagrangian generates self-coupled modes $\sum_k {\cal A}_k^4$ 
as well as couplings terms $\sum_{k,l}{\cal A}_k^2{\cal A}_l^2$ between the modes. This interacting 
field-theoretic problem cannot be solved any longer. However, 
since the normal mode dominates the dynamics at low energies closed to
the Euler instability, one can 
neglect the higher modes in this regime. Hence, we choose the ansatz 
$\phi(s,t) = {\cal A}_1(t) g_1(s)$ in the regime $F_0 \rightarrow F_c$ and restrict the 
discussion in the rest of our work to this regime. The so-far classical field theory can 
be quantized by introducing the canonically conjugate momentum ${\cal P} \equiv - i \hbar \partial / 
\partial {\cal A}_1$ and the 
time-dependent driving force can straightforwardly be included. 
Note that when the driving frequency is close to the fundamental
frequency of the beam, the fundamental mode will dominate also in
absence of a static longitudinal compression force. However, a
compression force helps to enhance the nonlinear effects which are in
the focus of this work. 
After all, an effective quantum mechanical time-dependent Hamiltonian results which 
describes the dynamics of a single quantum particle with ``coordinate'' ${\cal X}\equiv {\cal A}_1$ 
in a time-dependent anharmonic potential. It reads
\begin{equation}
H (t) = \frac{{\cal P}^2}{2m_{\rm eff}} + \frac{m_{\rm eff} \omega_1^2}{2}{\cal X}^2 + 
\frac{\alpha}{4} {\cal X}^4 + {\cal X} F(t) \label{hamsys}
\end{equation}
with the effective mass $m_{\rm eff} = 3 \sigma L / 8$ and the nonlinearity 
parameter $\alpha = (\pi/L)^4 F_c L (1+3 \epsilon)$. The classical
analogous system is the Duffing oscillator \cite{Nayfeh} 
(when, in addition, damping is included, see below). 
It shows a rich variety of features including
regular and chaotic motion. 
In this work, we focus on the parameter regime where only regular motion
occurs. For weak driving strengths, the response as a function of the 
driving frequency $\omega_{\rm ex}$ 
has the well-known form of the harmonic oscillator with the maximum at
$\omega_{\rm ex}=\omega_1$. 
For increasing
driving strength, the resonance grows and bends away from the $\omega_{\rm ex}=\omega_1$-axis 
towards larger frequencies (since $\alpha>0$). The locus of the maximal amplitudes is 
 often called the {\em backbone curve\/} \cite{Nayfeh}. 
 The corresponding nonlinear response of the quantum system shows
 clear signatures of sharp multi-phonon resonances whose line shapes
 will be calculated below. 

\subsection{Phenomenological model for damping\label{modelbath}}

In our approach, 
we do not intend to focus on the role 
of the microscopic damping mechanisms as this depends on the details
of the experimental device. Instead, we 
introduce damping phenomenologically in the standard way \cite{Weiss99}
by coupling the resonator Hamiltonian Eq.\ (\ref{hamsys}) to a bath 
of harmonic oscillators described by the standard Hamiltonian 
\begin{equation}
H_B=\frac{1}{2}\sum_j \frac{p_j^2}{m_j} + m_j \omega_j^2 \left( x_j- 
\frac{c_j}{m_j \omega_j^2} {\cal X} \right)^2\, ,  
\end{equation}
with the spectral density 
\begin{equation}
J(\omega)=\frac{\pi}{2}\sum_j\frac{c_j^2}{m_j\omega_j}\delta(\omega-\omega_j)=
m_{\rm eff}\gamma_s \omega_1^{1-s}\omega^s e^{-\omega/\omega_c}\, , 
\end{equation}
with  damping constant $\gamma_s$ and cut-off 
frequency $\omega_c$.  Our results discussed below 
 are valid for an Ohmic ($s=1, \gamma_1\equiv \gamma)$ as well as 
 for super-Ohmic ($s>1$) baths.  
Sub-Ohmic baths will not be considered here since the weak-coupling
assumption which allows the Markov approximation does not hold any
longer. Formally, the coefficients in the master equation would
diverge in the sub-Ohmic case, see Eq.\ (\ref{popnum}) below. 
The total Hamiltonian is $H_{\rm tot}(t)=H(t)+H_B$. 

To proceed, we scale $H_{\rm tot}(t)$ to dimensionless quantities 
such that the energies are in units of $\hbar \omega_1$ while the lengths are scaled in units of 
$x_0\equiv \sqrt{\frac{\hbar}{m_{\rm eff} \omega_1}}$. Put differently, we
formally set $m_{\rm eff}=\hbar=\omega_1=1$. 
The nonlinearity parameter $\alpha$ is scaled in units of 
$\alpha_0\equiv \hbar \omega_1/{\cal X}_0^4$, while the driving amplitudes are given in units of $f_0\equiv 
\hbar \omega_1 / x_0$. Moreover, we scale temperature in units of $T_0\equiv \hbar \omega_1 / k_B$ while the
damping strengths are measured with respect to $\omega_1$. 
\section{Coherent dynamics and rotating wave approximation (RWA)}
\label{coherent}
\begin{figure}[h]
\begin{center}
\vspace*{25mm}
\epsfig{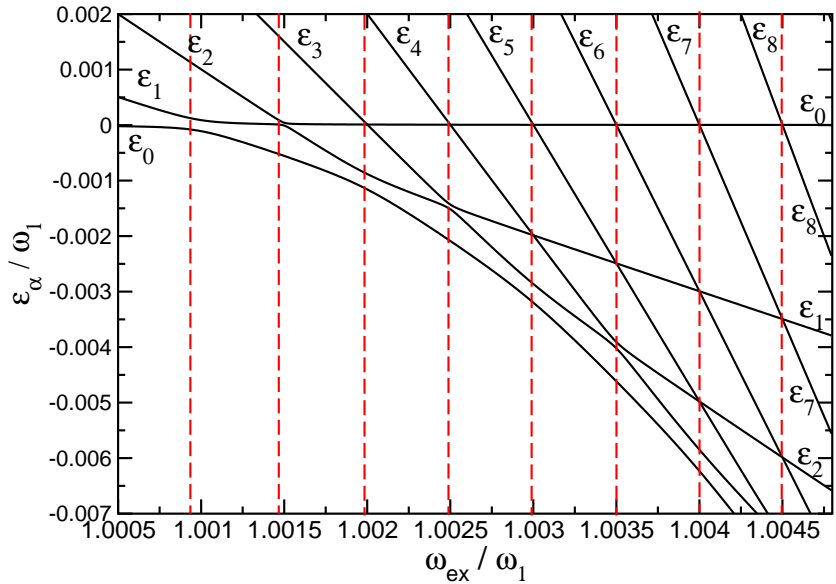}
\caption{Typical quasienergy spectrum $\varepsilon_\alpha$ 
for increasing  driving
frequency $\omega_{\rm ex}$ for the case $\nu=10^{-3}$ and
$f=10^{-4}$. The vertical dashed lines indicate the multiple avoided
level crossings for a fixed driving frequency. \label{fig.00}}
\end{center}
\end{figure}

Let us first consider the resonator dynamics without coupling to the bath. 
For convenience, we switch to a representation in terms of creation and annihilation 
operators $a$ and $a^\dagger$, such that ${\cal  X} = x_0 (a + a^\dagger)/\sqrt{2}$. 
Moreover, it is convenient to switch to the rotating frame by formally performing 
the canonical transformation $R=\exp{[-i\omega_{\rm ex} a^\dagger a t]}$. 
We  are interested in the nonlinear response of the resonator around
its fundamental frequency, i.e., for $\omega_{\rm ex} \approx \omega_1$, 
and will not consider the response at higher
harmonics. We further assume that  
the driving amplitude $f$ is not too large such that the nonlinear effects are 
small enough in order not to enter the chaotic regime. 
This suggests to use a rotating wave approximation 
(RWA) of the 
full system Hamiltonian $H(t)$ in Eq.\ (\ref{hamsys}) as the 
fast oscillating terms will be negligible around the fundamental
frequency for weak enough driving. 
By eliminating all the fast oscillating terms 
from the transformed Hamiltonian, one obtains  the Schr\"odinger 
equation in the rotating frame 
$\tilde{H} |\phi_\alpha\rangle = \varepsilon_\alpha 
|\phi_\alpha\rangle$ with the Hamiltonian in the RWA 
\begin{equation}
\tilde{H}=
%\tilde{\omega} \hat{n}+\frac{\nu}{2} \hat{n}(\hat{n}+1)+\frac{x_0}{2} f[a+a^\dagger e^{2i\omega_1t}+a e^{-2i\omega_1t}+ a^\dagger]+f_2[a e^{i[(\omega_2-\omega_1)t+\theta_{12}]}\nonumber\\
%&+&a^\dagger e^{i[(\omega_2+\omega_1)t+\theta_{12}]}+a e^{-i[(\omega_2+\omega_1)t+\theta_{12}]}+a^\dagger e^{-i[(\omega_2-\omega_1)t+\theta_{12}]}]\nonumber\\
\tilde{\omega} \hat{n}+\frac{\nu}{2} \hat{n}(\hat{n}+1)+f\left(a+a^\dagger\right) \, .
\label{htilde}
\end{equation}
Here, we have introduced  the detuning
 $\tilde{\omega}=\omega_1-\omega_{\rm ex}$, 
the nonlinearity parameter $\nu=3 \hbar \alpha /(4 \omega_1^2)$, 
$f=\tilde{f}(8 \hbar \omega_1)^{-1/2}$ and 
$\hat{n}=a^\dagger a$. In the static frame, an orthonormal basis (at
equal times) follows as 
\begin{equation}
|\varphi_\alpha(t)\rangle= e^{-i\omega_{\rm ex} a^\dagger a t}|\phi_\alpha\rangle \, .
\end{equation} 
The Hamiltonian (\ref{htilde}) has been studied in Refs.\ 
\cite{Larsen76,Dykman05}. 
The quasi-energy levels for a given number $N$ of phonons are pairwise degenerate, 
$\varepsilon_{N-n}=\varepsilon_n$ for $n\le N$, vanishing $f \rightarrow 0$ 
and $\tilde{\omega}=-\nu(N+1)/2$. For a finite  driving strength $f > 0$, the exact 
crossings turn then into avoided crossings which is a signature of multiphonon 
transitions \cite{Peano04,Dykman05}. A typical quasienergy spectrum is
shown in Fig.\  \ref{fig.00} for the parameters $\nu=10^{-3}$ and
$f=10^{-4}$. The dashed vertical lines indicate the multiple avoided
level crossings which occur all for the same driving frequency. 
For $|\varepsilon|=|2f/[\nu(N+1)]|\ll 1$, 
each pair of degenerate levels interacts only weakly with the 
other levels, and act effectively like a two-level 
Rabi system \cite{Larsen76}. The Rabi frequency is related to the 
minimal splitting of the levels and is calculated perturbatively 
with $\varepsilon$ as a small parameter in the following Section. 

\subsection{Van Vleck perturbation theory}

Let us therefore consider  the multiphonon resonance 
at $\tilde{\omega}=-\nu(N+1)/2$. In addition, we are interested in the
response around the resonance and therefore introduce the 
small deviation $\Delta$. We formally rewrite $\tilde{H}$ as
\begin{equation}
\tilde{H}=\frac{\nu(N+1)}{2}\left[-(1+\Delta)\hat{n}
+\frac{\hat{n}+1}{N+1}\hat{n}+\varepsilon(a+a^\dagger)\right]
\,. 
\end{equation}
Let us then first discuss the dynamics at resonance ($\Delta=0$).  
%In the following we will consider a finite $\Delta$ in order to describe the system 
%near the resonance. Let us first focus on the resonant behavior, corresponding to $\Delta=0$.
 %In order to simplify the notation, we rescale $\tilde{H}$ by
 %$\nu(N+1)/2$. 
 We divide 
 it in the unperturbed part $H_0$ and the perturbation $\varepsilon V$
 according to 
\begin{equation}
H_0=\frac{\nu(N+1)}{2}\left[ 
-\hat{n}+\frac{\hat{n}+1}{N+1}\hat{n}\right] 
\,,\quad V=\frac{\nu(N+1)}{2}\left[a+a^\dagger \right]\,,
\label{ham0}
\end{equation} 
respectively. The unperturbed Hamiltonian is diagonal and  near the resonance its spectrum 
is divided in well separated groups of nearly degenerate quasienergy eigenvalues. 
An appropriate perturbative method to diagonalize this type of Hamiltonian
 is the Van Vleck perturbation theory \cite{Cohbook,Sha80,Goorden05}. 
 It defines a unitary 
 transformation yielding  
 the Hamiltonian $\tilde{H}$ in an effective block diagonal form. The effective 
 Hamiltonian has the same eigenvalues as the original one, 
 with the quasi-degenerate eigenvalues in a common block. The effective 
 Hamiltonian can be written as 
 \begin{equation}
\tilde{H}'=e^{iS}\tilde{H} e^{-iS}\,. 
\label{VanVlecktrans}
\end{equation}
 In our case, each block is a two by two matrix
  corresponding to a subspace
  formed by a couple of quasienergy states forming an anticrossing. 
  Let us consider the effective Hamiltonian $H'_n$
   corresponding to the involved levels $|n\rangle$ %$|\phi_n\rangle$ 
   and $|N-n\rangle$, being eigenstates of the harmonic oscillator. 
     %$|\phi_{N-n}\rangle$.    
   The degeneracy in the corresponding 
   block is lifted at order $N-2n$ in    
   Van Vleck perturbation theory. The block Hamiltonian then reads   
 \begin{equation}
 \tilde{H}'_n=
 \left(
 \begin{array}{cc}
 \frac{\nu}{2}n(n-N)%+\sum_{m=1}^\infty\varepsilon^m C_{11,m}
 & \varepsilon^{N-2n} C_{12,N-2n}\\
 \varepsilon^{N-2n} C_{12,N-2n} 
 &\frac{\nu}{2}n(n-N)\\
 \end{array}
 \right) \, , \label{loworderham}
 \end{equation}
where 
 \begin{equation}
 C_{12,N-2n}=(N+1)^{N-2n}\frac{\nu}{2}
 \frac{\sqrt{(N-n)!}}{\sqrt{n!}(N-2n-1)!^{2}} \, .
 \end{equation}
This is the lowest order of the perturbed
 Hamiltonian which allows to calculate the corresponding zero-th order
 eigenstates.  
% In the Appendix we will prove that the lowest order contribution to the off-diagonal elements is
% $C_{12,N-2n}$ and that the two levels are still degenerate up to fifth order in perturbation theory.
% Corrections to the diagonal elements of  $m-$th order with $6\le m<N-2n$ could lift the degeneracy between the two levels. This would have the physical meaning of a  shift of the resonance position of  $m-$th order. 
% %=(N+1)^{N-2n-1}\sqrt{(N-n)!}/(\sqrt{n!}(N-2n-1)!^2)$ 
%If such a shifting occurs, in order to study the system at resonance one should determine $\Delta$ self-consistently, by formally expanding $\Delta$ in terms of $\varepsilon$  
%as $\Delta=\sum_{i=m}^N \varepsilon^i \delta_i$.  
%One should  then   fix the $\delta_i$'s so that the levels remain degenerate  
%up to $(N-2n-1)-$th order. In any case, the degeneracy is only lifted at 
% $(N-2n)-$th. \\
% 
 By diagonalizing  $\tilde{H}'_n$ in Eq.\  (\ref{loworderham}), one finds  the  minimal 
 splitting for the $N-$phonon transition as 
 \begin{eqnarray}
 \Omega_{N,n} &=&|2\varepsilon^{N-2n} C_{12,N-2n}|\nonumber\\
 &=&2 f
\left(\frac{2
f}{\nu}\right)^{N-2n-1}\frac{\sqrt{(N-n)!}}{\sqrt{n!}(N-2n-1)!^{2}} \,
.
 \label{Rabi}
 \end{eqnarray}
For the case away from resonance, we consider a detuning
$\Delta=\varepsilon^N \delta$.  Within the Van Vleck technique, only the
zero-th block is influenced according to  
 \begin{equation}
 \tilde{H}'_0=
 \left(
 \begin{array}{cc}
 0 & \varepsilon^{N} C_{12,N}\\
 \varepsilon^{N} C_{12,N} &-\frac{\nu(N+1)}{2}\varepsilon^N N  \delta \\
 \end{array}
 \right) \, ,
 \label{vanvleckham}
 \end{equation}
the other blocks given in Eq.\  (\ref{loworderham}) for $n\ne 0$ 
are not influenced by this higher-order correction.
%Let us now consider a detuning 
%Assuming that the resonance positions are not shifted (this has been proved in the 
%appendix for $N\le6$)
%at resonance and zero order in perturbation theory
The eigenvectors for the Hamiltonian $\tilde{H}$ at zero-th order  are 
obtained by diagonalizing $\tilde{H}_n^\prime$ in Eq.\  (\ref{loworderham}).
One finds $|\phi_{n}\rangle=|n\rangle$ for $n\ge N+1$ or 
 $|\phi_{n}\rangle=(|n\rangle + |N-n \rangle)/\sqrt{2}$ and 
  $|\phi_{N-n}\rangle=(|n\rangle - |N-n \rangle)/\sqrt{2}$
for $0< n <N/2$ and $|\phi_{N/2}\rangle = |N/2 \rangle$ if $N$ is even. 
Moreover, 
\begin{eqnarray}
|\phi_{0}\rangle & = & \cos \frac{\theta}{2} |0 \rangle - \sin \frac{\theta}{2} 
|N \rangle \, , \nonumber \\
|\phi_{N}\rangle & = & \sin \frac{\theta}{2} |0 \rangle + \cos \frac{\theta}{2} 
|N \rangle \, , 
\label{eigenvectors}
\end{eqnarray}
where we have introduced the angle $\theta$ via $\tan \theta =
-2\Omega_{N,0}/[\nu (N+1) N\Delta ]$. 

\section{Dissipative dynamics in presence of the bath
\label{dissipative}}

Having discussed the coherent dynamics,  we include now 
the influence of the harmonic bath coupled to the driven system. We
therefore assume that the coupling is weak enough such that the
standard Markovian master equation 
\begin{equation}
\frac{d}{dt}\varrho=-i[H(t),\varrho]+{\cal L} \varrho\,
\end{equation}
for the reduced density operator $\rho(t)$ can be applied. 
The influence of the bath enters in the superoperator 
\begin{equation}
{\cal L}\varrho=-[ {\cal X},[P(t),\varrho]_+]-
[{\cal X},[Q(t),\varrho]]
\label{supop}
\end{equation}
with the correlators
\begin{equation}
P(t)=\frac{i}{2} \int_0^\infty d\tau \gamma(\tau)U^\dagger(t-\tau,t){\cal P}
U(t-\tau,t)\, 
\end{equation}
and 
\begin{equation}
Q(t)=\int_0^\infty d\tau K(\tau)U^\dagger(t-\tau,t){\cal X}
U(t-\tau,t)\, .
\end{equation}
The kernels are given by 
\begin{eqnarray}
\gamma (\tau) & = &  \frac{2}{\pi m_{\rm eff}} \int_0^\infty 
\frac{J(\omega)}{\omega} \cos \omega \tau \, , \nonumber \\
K(\tau) & = & \frac{1}{\pi} \int_0^\infty J(\omega) \coth \left( 
\frac{\hbar
\omega}{2 k_B T} \right) \cos \omega \tau \, ,
\end{eqnarray}
where $T$ is the environment temperature. Moreover, $U(t,t')= {\cal T} 
\exp (i\int_{t'}^t H(t) dt)$ is the propagator with the time order
operator ${\cal T}$. Next, we project the density matrix on the orthonormal set
$|\varphi_\alpha (t)\rangle = 
\exp{[-i\omega_{\rm ex} a^\dagger a t]}|\phi_\alpha\rangle$, such that 
the matrix elements read
\[
\varrho_{\alpha\beta}(t)= \langle\phi_\alpha|e^{i\omega_{\rm ex} a^\dagger a t}\varrho(t)
e^{-i\omega_{\rm ex} a^\dagger a t}|\phi_\beta\rangle \, .
\]  
Performing the derivative one obtains 
\begin{eqnarray}
\dot{\varrho}_{\alpha\beta}(t)&=&-i\langle\phi_\alpha|
e^{i\omega_{\rm ex} a^\dagger a t}\left[i
\stackrel{\leftarrow}{
\frac{d}{dt}}\varrho+[H(t),\varrho]+i{\cal L} \varrho+ 
\varrho i\frac{d}{dt}\right]
e^{-i\omega_{\rm ex} a^\dagger a t}|\phi_\beta\rangle \nonumber \\
&\simeq&-i(\varepsilon_\alpha-\varepsilon_\beta)\varrho_{\alpha\beta}(t)+
\langle\phi_\alpha(t)|e^{i\omega_{\rm ex} a^\dagger a t}\,{\cal L}\varrho\, 
e^{-i\omega_{\rm ex} a^\dagger a t}|\phi_\beta(t)\rangle\,.
\label{MME}
\end{eqnarray}
%
%in the second line we have used twice Eqs.\ (\ref{htilde}) and 
%(\ref{eigenvector}). 
%
For the dissipative term, we need to  compute
\begin{eqnarray}
X_{\alpha\beta}(t)
&=&\langle\phi_\alpha|e^{i\omega_{\rm ex} a^\dagger a t}
{\cal X} \,
e^{-i\omega_{\rm ex} a^\dagger a t}|\phi_\beta\rangle\nonumber\\
%&=&x_0e^{-i\omega_{\rm ex} t}\langle\phi_\alpha|a|\phi_\beta\rangle
%+x_0e^{i\omega_{\rm ex} t}\langle\phi_\alpha|a^\dagger|\phi_\beta\rangle\nonumber\\
&=&\left( e^{-i\omega_{\rm ex} t} X_{\alpha\beta,+1}
+e^{i\omega_{\rm ex} t} X_{\beta\alpha,-1} \right) \,,
\label{pos}
\end{eqnarray}
with $X_{\alpha\beta,+1}=X^*_{\beta \alpha,-1} = x_0 
\langle\phi_\alpha|a|\phi_\beta\rangle/\sqrt{2}$ 
being the matrix element of the  
destruction operator in the rotating frame. 
We need, moreover, 
\begin{eqnarray}
Q_{\alpha\beta}(t)&=&\int_0^\infty d\tau K(\tau)\langle\phi_\alpha
|e^{i\omega_{\rm ex} a^\dagger a t}U^\dagger(t-\tau,t){\cal X} U(t-\tau,t)
e^{-i\omega_{\rm ex} a^\dagger a t}|\phi_\beta\rangle\nonumber\\
&=&\int_0^\infty d\tau K(\tau)e^{-i(\varepsilon_\alpha-\varepsilon_\beta)\tau} 
\langle\phi_\alpha|e^{i\omega_{\rm ex} a^\dagger a (t-\tau)} {\cal X} 
e^{-i\omega_{\rm ex} a^\dagger a (t-\tau)}|\phi_\beta\rangle\\
%&=&\int_0^\infty d\tau K(\tau)e^{-i(\varepsilon_\alpha-\varepsilon_\beta)\tau}X(t-\tau)\\
&=&e^{-i\omega_{\rm ex} t}\left[\int_0^\infty d\tau K(\tau)
e^{-i(\varepsilon_\alpha-\varepsilon_\beta-\omega_{\rm ex})\tau}\right]
X_{\alpha\beta,+1}\nonumber\\
& & +e^{i\omega_{\rm ex} t}\left[\int_0^\infty d\tau K(\tau)
e^{-i(\varepsilon_\alpha-\varepsilon_\beta
+\omega_{\rm ex})\tau}\right] 
X_{\alpha\beta,-1} \, .
\label{Q}
\end{eqnarray}
In an analogous way, we have
\begin{eqnarray}
P_{\alpha\beta}(t) & = & \frac{i}{2} \int_0^\infty d\tau \gamma(\tau)
e^{-i(\varepsilon_\alpha-\varepsilon_\beta)\tau} 
\langle\phi_\alpha|e^{i\omega_{\rm ex} a^\dagger a (t-\tau)} {\cal P} 
e^{-i\omega_{\rm ex} a^\dagger a (t-\tau)}|\phi_\beta\rangle
\nonumber \\
&= & -\frac{m_{\rm eff}}{2} \int_0^\infty d\tau \gamma(\tau)
e^{-i(\varepsilon_\alpha-\varepsilon_\beta)\tau} \nonumber \\
&& \times 
\left(
\langle\phi_\alpha|e^{i\omega_{\rm ex} a^\dagger a
(t-\tau)}[\pm i\stackrel{\leftrightarrow}{\frac{d}{dt}}+H(t) ,
{\cal X}]
e^{-i\omega_{\rm ex} a^\dagger a (t-\tau)}|\phi_\beta\rangle\right.\nonumber\\
&&\left.-i\frac{d}{dt}\langle\phi_\alpha|
e^{i\omega_{\rm ex} a^\dagger a (t-\tau)} {\cal X}\,
e^{-i\omega_{\rm ex} a^\dagger a (t-\tau)}|\phi_\beta\rangle\right)\nonumber\\
&\simeq&-\frac{m_{\rm eff}}{2} \left(\varepsilon_\alpha-\varepsilon_\beta
-i\frac{d}{dt}\right)
\left[\int_0^\infty d\tau \gamma(\tau)
e^{-i(\varepsilon_\alpha-\varepsilon_\beta)\tau}X_{\alpha
\beta}(t-\tau)\right] \, . \nonumber\\
\label{P}
\end{eqnarray}
Here, we have defined the time derivative  
$\pm \stackrel{\leftrightarrow}{\frac{d}{dt}}$  where the positive
(negative) sign belongs to the left (right) direction. In the second line,
 we have used  the canonical relation ${\cal P}/m_{\rm eff}=-i[{\cal
 X}, H (t)]$.  We can 
now compute the matrix elements of the operators involved in the dissipative part
  in Eq.\ (\ref{MME}) and find for the terms in Eq.\ 
  (\ref{supop})
\begin{equation}
\left(P+Q\right)_{\alpha\beta}=
\left(e^{-i\omega_{\rm ex} t}N_{\alpha\beta,-1}X_{\alpha\beta,+1}
+e^{i\omega_{\rm ex} t}N_{\alpha\beta,+1}X_{\alpha\beta,-1}\right)
\end{equation}
and
\begin{equation}
\left(P-Q\right)_{\alpha\beta}=-
\left(e^{-i\omega_{\rm ex} t}N_{\beta\alpha,+1}X_{\alpha\beta,+1}
+e^{i\omega_{\rm ex} t}N_{\alpha\beta,-1}X_{\alpha\beta,-1}\right)\,.
\end{equation}
Here,  $N_{\alpha\beta,\pm 1}$ are defined as
\begin{equation}
N_{\alpha\beta,\pm1}=N(\varepsilon_\alpha-\varepsilon_\beta\pm\omega_{\rm ex})\quad
N(\varepsilon)=J(|\varepsilon|)[n_{th}(|\varepsilon|)+\theta(-\varepsilon)]\,, 
\label{popnum}
\end{equation}
in terms of the bath density of states $J(|\varepsilon|)$, 
the bosonic thermal occupation number
\begin{equation}
n_{th}(\varepsilon)=\frac{1}{2}\left[\coth{\left(\frac{\varepsilon}{2k_BT}\right)}
-1\right]\,
\end{equation}
and the Heaviside function $\theta(x)$.% indicating spontaneous emission.

Eq.\  (\ref{popnum}) illustrates 
why we have to restrict  to (super-)Ohmic baths, since
$N(\varepsilon)$ 
would diverge for $s<1$ at low energies. 
During the calculation, the $\tau$-integration in the double integrals 
in Eqs.\ (\ref{P})  and (\ref{Q}) has 
been evaluated by using the representation 
$\int_0^\infty\,d\tau \exp{(i\omega\tau)}=\pi\delta(\omega)+i{\cal
P}_p(1/\omega)$, 
where ${\cal P}_p$ denotes the principal part. 
The contributions of the principal part result in quasienergy shifts of the 
order of $\gamma_s$ which are the so-called Lamb shifts. As usual, these 
have also been neglected here. 

The ingredients can now be put together to obtain the Markovian 
 master equation in the static frame as 
\begin{eqnarray}
\dot{\varrho}_{\alpha\beta}(t) = 
-i(\varepsilon_\alpha-\varepsilon_\beta)\varrho_{\alpha\beta}(t)
\nonumber \\
+ \sum_{\alpha' \beta'} \sum_{n,n'=\pm 1} 
e^{-i(n+n')\omega_{\rm ex} t} \left[ (N_{\alpha \alpha',-n}+N_{\beta \beta',n'})
X_{\alpha \alpha',n} \rho_{\alpha \beta'}X_{\beta' \beta,n'} \right. \nonumber
\\
\left. 
- N_{\beta' \alpha',-n'}X_{\alpha \beta',n}X_{\beta' \alpha',n'}
\rho_{\alpha' \beta}
-N_{ \alpha' \beta',n'}\rho_{\alpha \beta'}X_{ \beta' \alpha',n'}
X_{\alpha' \beta,n} \right] \, .
\label{fullME}
\end{eqnarray}
Next, we perform a `moderate rotating wave approximation' 
consisting in averaging the time-dependent terms in the 
bath part over the driving period $T_{\omega_{\rm ex}} = 2\pi / \omega_{\rm ex}$. 
This is consistent with the assumption of weak coupling which assumes
that dissipative effects on the dynamics are noticeable only on a time
scale much larger than $T_{\omega_{\rm ex}}$. 
 Under this approximation, the master equation becomes 
\begin{eqnarray}
\dot{\varrho}_{\alpha\beta}(t) & = & \sum_{\alpha' \beta'}{\cal S}_{\alpha\beta,\alpha'\beta'}
\varrho_{\alpha'\beta'}(t) \nonumber \\
& = &\sum_{\alpha' \beta'}\left[-i(\varepsilon_\alpha-\varepsilon_\beta)\delta_{\alpha\alpha'}
\delta_{\beta\beta'}+{\cal L}_{\alpha\beta,\alpha'\beta'}\right)]
\varrho_{\alpha'\beta'}(t)
\label{ME}
\end{eqnarray}
with the dissipative transition rates
\begin{eqnarray}
{\cal L}_{\alpha\beta,\alpha'\beta'} & = & 
\sum_{n=\pm1} (N_{ \alpha \alpha',-n}+N_{ \beta \beta',-n}) X_{\alpha \alpha',n}
X_{\beta' \beta,-n} \nonumber \\
& & -\delta_{\alpha \alpha'} \sum_{\alpha''; n=\pm 1}
N_{ \alpha'' \beta',-n}X_{ \beta' \alpha'',-n}
X_{\alpha'' \beta,n} \nonumber \\
& & -\delta_{\beta \beta'} \sum_{\beta''; n=\pm 1}
N_{ \beta'' \alpha',-n}X_{ \alpha \beta'',-n}
X_{\beta''\alpha',n} \, . 
\label{MErate}
\end{eqnarray}
It is instructive to compare this master equation 
to the one in Ref.\ \cite{Kohler97}, given in terms of the 
full Floquet quasienergy states.  The key difference here is 
that the density matrix is projected onto the approximate eigenvectors 
$\exp{(-i\omega_{\rm ex} a^\dagger a t)}|\phi_\alpha\rangle$ rather than onto 
the exact Floquet solutions. 
 As a consequence of the RWA,
the sums in Eq.\ (\ref{MErate}) only include the $n=\pm 1$ terms
indicating that only one-step transitions are possible where 
$n=-1$ refers to emission and $n=+1$ to absorption.  
Being consistent with the RWA, we can assume that 
$|\nu|,|f|,|\omega_{\rm ex}-\omega_1|\ll 
\omega_1$ which yields to $|\varepsilon_\alpha-\varepsilon_\beta|\ll\omega_{\rm ex}$. 
Hence,  $N_{\alpha \beta, + 1}$ is the product of the bath
density of states and the bosonic occupation number at temperature
$T$. This corresponds to the thermally activated absorption of a phonon 
from the bath. 
On the other hand,  $N_{\alpha \beta, - 1}$ given in Eq.\
(\ref{popnum}) contains the
temperature-independent term $..+J(\omega_{\rm ex})$ describing 
spontaneous emission. 

\section{Observable for the nonlinear response \label{observables}}

 Assuming an ergodic dynamics of the full system, or equivalently that there 
 is just one eigenvector $\varrho^\infty$ of the superoperator 
${\cal S}$ in Eq.\ (\ref{ME}), corresponding to a vanishing eigenvalue, and 
that all the other eigenvalues have negative real part,  the asymptotic 
solution of Eq.\ (\ref{ME}) is
\begin{equation}
\lim_{t\rightarrow\infty}\varrho(t)=\varrho^\infty \, .
\label{assol}
\end{equation}
To simplify notation, we omit in the following the superscript 
$\infty$ but only refer to the stationary state $\varrho_{\alpha
\beta} \equiv \varrho_{\alpha
\beta}^\infty$. 

We are interested in the mean value of the position operator in the
stationary state according to 
\begin{equation}
\langle {\cal X}  \rangle_t=tr(\varrho {\cal X}) 
 = \sum_{\alpha \beta}\varrho_{\alpha\beta}X_{\beta\alpha}(t)\, .
 \end{equation}
Using Eq.\ (\ref{pos}) yields 
%\begin{equation}
$\langle {\cal X}  \rangle=A \cos{(\omega_{\rm ex} t+\varphi )}$, 
% \end{equation}
with the oscillation amplitude 
\begin{equation}
A= 2
|\sum_{\alpha\beta}\varrho_{\alpha\beta}X_{\beta\alpha,+1}|\, ,
\label{nonlinearres}
 \end{equation}
 and the phase shift
\begin{equation}
\varphi
=\pi \theta \left(-{\rm Re}\left[\sum_{\alpha\beta}
\varrho_{\alpha\beta}
X_{\beta\alpha,+1}\right]\right)
+\arctan{\left[\frac{{\rm Im}\left[\sum_{\alpha\beta}\varrho_{\alpha\beta}
X_{\beta\alpha,+1}\right]}{{\rm Re}\left[\sum_{\alpha\beta}
\varrho_{\alpha\beta}
X_{\beta\alpha,+1}\right]}\right]} \, , 
 \end{equation} 
with $\theta$ being the Heaviside function. 
\section{Analytical solution for the lineshape of the multiphonon
resonance  in the perturbative regime \label{sol}}
When the driving frequency $\omega_{\rm ex}$ is varied, the amplitude
$A$ shows characteristic multi-phonon resonances at those values for
which the quasienergy levels form avoided level crossings
\cite{Peano04}. While in Ref.\ \cite{Peano04} these resonances have
been studied numerically, it is the central result of this work to
calculate their line shape analytically by solving 
the corresponding master equation in the Van Vleck perturbative regime.  

Within the limit of validity of the RWA, i.e., $|\nu|,|f|,|\omega_{\rm ex}-\omega_1|\ll 
\omega_1$, we have $|\varepsilon_\alpha-\varepsilon_\beta|\ll\omega_{\rm ex}$. 
In the regime of low temperature $k_B T\ll \omega_{\rm ex}$, it
follows from Eq.\ (\ref{popnum})  
that $N_{\alpha \beta, -1}\simeq J(\omega_{\rm ex})$ 
and $N_{\alpha \beta, 1}\simeq 0$ entering in the transition rates 
in Eq.\ (\ref{MErate}). 
This approximation corresponds to consider spontaneous emission only 
and yields the dissipative transition rates
\begin{eqnarray}
{\cal L}_{\alpha\beta,\alpha'\beta'} & = & 
 \frac{\gamma_s}{2}\left(\frac{\omega_{\rm ex}}{\omega_1}\right)^s\left( 2 A_{\alpha \alpha'}
A_{\beta \beta'} 
-\delta_{\alpha \alpha'} \sum_{\alpha''}
A_{\alpha'' \beta'}
A_{\alpha'' \beta}\right. \nonumber \\
& &
 \left. -  \delta_{\beta \beta'} \sum_{\beta''}
A_{\beta'' \alpha}
A_{\beta''\alpha'}\right) \, .
\end{eqnarray}
Here, we have defined $A_{\alpha \beta}\equiv \langle \phi_\alpha | a
| \phi_\beta \rangle$.  Note that it is consistent with the previous
approximation to set $\omega_{\rm ex}/\omega_1\approx 1$. Hence, all
the following results are valid for Ohmic as well as super-Ohmic
baths.   

In the following we will  use this simplified transition rates to solve 
the master equation near the multiple multiphonon resonances. The 
transition between the groundstate and
the $N$-phonon state is the narrowest. Hence, it will be
affected first when a finite coupling to the bath is considered.  
In particular, it is interesting to consider the
case when the damping
constant  $\gamma_s$ is larger than the minimal splitting 
$\Omega_{N0}$ between the two quasienergy states but smaller than all
the minimal splittings of the other, i.e., 
$\Omega_{N0}<\gamma_s\ll\Omega_{Nn}$ for $n\ge1$.  
In this case, we can assume a partial secular approximation: We set all the 
off-diagonal elements to zero except for $\varrho_{0N}$ and 
$\varrho_{N0}=\varrho^*_{0N}$. In this regime 
the stationary solutions are determined by the conditions
\begin{eqnarray}
0&=&\sum_{\beta}{\cal L}_{\alpha\alpha,\beta\beta}\varrho_{\beta\beta}
+{\cal L}_{\alpha\alpha,0 N} \, \, 2 \, \, {\rm Re}(\varrho_{0 N}) \,
,
\nonumber \\
0&=&-i(\varepsilon_0-\varepsilon_N)\varrho_{0N}+
\sum_{\alpha}{\cal L}_{0N,\alpha\alpha}\rho_{\alpha \alpha}+{\cal L}_{0N,0N}\varrho_{0N}
+{\cal L}_{0N,N0}\varrho^*_{0N}\,.
\label{psecular}
\end{eqnarray}

For very weak damping, i.e., when $\gamma_s$ is smaller than all 
minimal splittings ($\gamma_s\ll\Omega_{Nn}$), 
the off-diagonal elements of the density matrix are negligibly small and 
can be set  to zero. Within this approximation, the stationary solution for the 
density matrix is determined by the simple kinetic equation
\begin{equation}
0=\sum_{\beta}{\cal L}_{\alpha\alpha,\beta\beta}\varrho_{\beta\beta}\,.
\label{secular}
\end{equation}
In this regime, a very simple physical picture arises. The bath causes 
transitions between different quasienergy states, but here, 
the transition rates are independent from the quasienergies. It is 
instructive to express  the quasienergy solutions in terms of the harmonic oscillator 
(HO) solutions as $|\phi_\alpha\rangle=\sum_n c_{\alpha n}|n\rangle$ with
some coefficients $c_{\alpha n}$. The transition 
rates between two quasienergy states then read 
\begin{equation}
{\cal L_{\alpha\alpha,\beta\beta}}=\gamma_s 
|\langle\phi_\alpha | a
| \phi_\beta \rangle|^2=\gamma_s \sum_n 
(n+1) |c_{\alpha n}|^2|c_{\beta n+1}|^2 \, .
\label{transitionrate}
\end{equation}
This formula  illustrates simple selection rules in
this low-temperature regime: Only those components of the two
different quasienergy states contribute to the transition rate whose energy  
differs by one energy quantum  ($n \leftrightarrow n+1$).   
\subsection{One-phonon resonance vs.\  antiresonance 
\label{onephonon}}
Before we consider the general multiphonon case, we first elaborate on
the one-phonon resonance. This, in particular, allows to make the
connection to the standard linear response of a driven damped
harmonic oscillator which is resonant at the frequency $\omega_1+\nu$. 
We will illustrate the mechanism how this resonant
behavior is turned into an antiresonant behavior when the damping is
reduced (and the driving amplitude $f$ is kept fixed).  

The corresponding effective 
Hamiltonian $\tilde{H}_0^\prime$ follows from Eq.\  
(\ref{vanvleckham}) and is readily diagonalized by the quasienergy states 
$|\phi_0 \rangle$ and  $|\phi_1 \rangle$ which are of zero-th order 
in $\varepsilon$ and which are given in Eq.\  
(\ref{eigenvectors}). The master equation (\ref{psecular}) can be
straightforwardly solved in terms of the rates ${\cal L}_{\alpha
\beta, \alpha' \beta'}$ for which one needs the ingredients 
$A_{00}=-A_{11}=\sin(\theta/2)\cos(\theta/2), 
A_{01}=\cos^2(\theta/2)$ and $A_{10}=-\sin^2 (\theta/2)$.  
The general solution  follows as 
\begin{eqnarray}
\rho_{00} & = & \nonumber \\
& & \mbox{\hspace{-19ex}}
\frac{- {\cal L}_{00,11} [{\cal L}_{01,01}^2-{\cal
L}_{01,10}^2 + \Omega^2(\Delta)] + 2 
{\cal L}_{00,01}{\cal L}_{01,11}({\cal L}_{01,01}-{\cal L}_{01,10})
}{
({\cal L}_{00,00}-{\cal L}_{00,11}) [{\cal L}_{01,01}^2-{\cal
L}_{01,10}^2 + \Omega^2(\Delta)] - 2{\cal L}_{00,01}
({\cal L}_{01,00}-{\cal L}_{01,11})  ({\cal L}_{01,01}-{\cal
L}_{01,10})  
} \, , \nonumber \\
{\rm Re  } \rho_{01} & = & 
 \frac{-({\cal L}_{01,01}-{\cal L}_{01,10}) 
 [ {\cal L}_{01,11} + ({\cal L}_{01,00}-{\cal L}_{01,11}) 
 \rho_{00}]}
{
 {\cal L}_{01,01}^2-{\cal
L}_{01,10}^2 + \Omega^2(\Delta) }\, ,\nonumber \\
{\rm Im  } \rho_{01} & = & 
 \frac{\Omega(\Delta)}
{ {\cal L}_{01,01}-{\cal
L}_{01,10}} {\rm Re  } \rho_{01}  \, , 
\end{eqnarray}
where $\Omega(\Delta)=\varepsilon_0 - \varepsilon_1$. 

\begin{figure}[h]
\begin{center}
\vspace*{25mm}
\epsfig{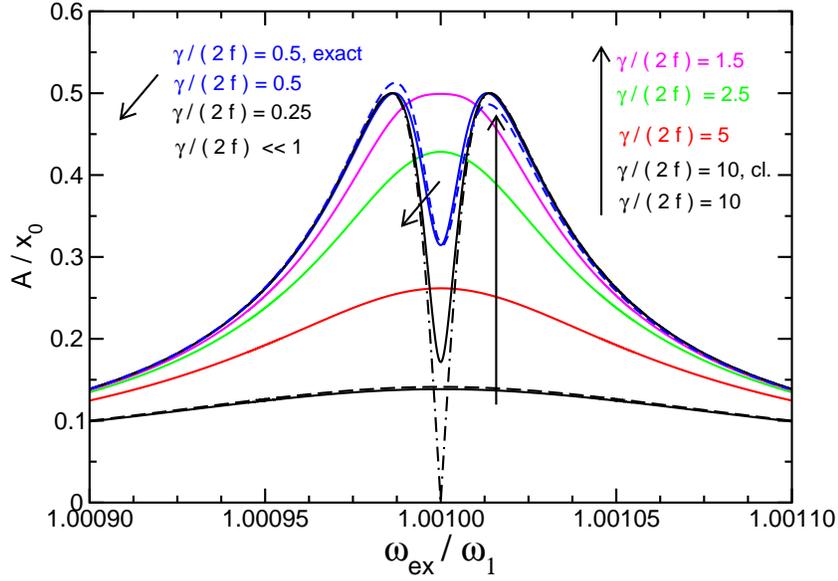}
\caption{Nonlinear response of the nanoresonator at the one-phonon
resonance $N=1$, where $\omega_{\rm ex} = \omega_1 + \nu$ for the
parameters $f=10^{-5}$ and $\nu=10^{-3}$ (in dimensionless units). 
The transition from a resonant  
behavior for large damping ($\gamma / (2f) = 10$) to an antiresonant 
 behavior at small damping ($\gamma / (2f) \ll  1$)is clearly visible. 
 The resonant line shape
 is a Lorentzian and coincides with the linear response of a harmonic
 oscillator at frequency $\omega_1+\nu$ (see black dashed line for 
 $\gamma / (2f) = 10$). Also shown is the 
 limit of $\gamma / (2 f) \ll 1$ (black dotted-dashed line) yielding a
 cusp-like  lineshape.   
 Note that we also depict   the solution from the exact master equation
for the case
$\gamma / (2f) = 0.5$ (blue dashed line). 
 \label{fig.1}}
\end{center}
\end{figure}

In the following, we calculate the amplitude $A$  according to Eq.\  
(\ref{nonlinearres}) to zero-th order in $\varepsilon$. 
In Fig.\  
\ref{fig.1}, we show the nonlinear response for the 
parameter set (in dimensionless units) 
$f=10^{-5}$ and $\nu=10^{-3}$. Moreover, the one-phonon resonance condition
reads $\omega_{\rm ex}=\omega_1+\nu$. The transition from the resonant
to antiresonant behavior depends on the ratio $\gamma/\Omega_{10} = 
\gamma/(2f)$. 
For the case of  stronger
damping $\gamma/(2f)=10$, we find that the
response shows a resonant behavior with a Lorentzian form similar to
the response of a damped
linear oscillator. In fact, the corresponding standard classical result is also
shown in Fig.\ \ref{fig.1} (black dashed line). 
The only effect of the nonlinearity to lowest order
perturbation theory  is to shift the
resonance frequency by the nonlinearity parameter $\nu$. 
The resonant behavior turns into an antiresonant one if the damping
constant is decreased to smaller values. A cusp-like line profile
arises in the limit of very weak damping when  
the damping strength is smaller than the minimal splitting, i.e.,  
$\gamma / (2 f) \ll 1$. Then, the response follows from the
master equation  (\ref{secular}) as 
\begin{equation}
A = x_0 \sqrt{2} \left|\sin \frac{\theta}{2}\cos \frac{\theta}{2} 
\right| 
\left| \frac{\sin^4\frac{\theta}{2}-\cos^4\frac{\theta}{2}}
{\sin^4\frac{\theta}{2}+\cos^4\frac{\theta}{2}}\right| \, .
\end{equation}
This antiresonance lineshape is also shown in Fig.\  \ref{fig.1} (see
dotted-dashed line). At resonance $\Delta=0$, we have an equal population of
the quasienergy states: $\rho_{00}=\rho_{11}=1/2$ and both add up 
to a vanishing oscillation amplitude $A$ since $A_{00}=-A_{11}$. 
Note that we show also the solution from the exact master equation
containing all orders in $\epsilon$, for the case
$\gamma / (2f) = 0.5$ and $s=1$ (blue dashed line in Fig.\  \ref{fig.1}), in
order to verify the validity of our perturbative treatment.  

\subsection{Multiphonon resonance vs.\ antiresonance 
\label{multiphonon}}
In this subsection we want to investigate the multiple multiphonon
resonances $N>1$. In order to 
illustrate the physics, we start with the simplest case at resonance
and within the secular approximation. 

\subsubsection{Secular approximation at resonance}

The zero-th order quasienergy solutions are given in terms of the 
eigenstates of the harmonic 
oscillator in Eq.\ (\ref{eigenvectors})
 with $\theta=\pi/4$. Then, $|n\rangle$ and 
 $|N-n\rangle$ ($n<N/2$) form a pair of quasienergy solutions. For N
 odd, there are $(N+1)/2$ pairs. For N even, there are $N/2$ pairs 
 whereas the state $|\phi_{N/2}\rangle=|N/2\rangle$ remains sole. 
 Within the secular approximation, we can describe the 
 dynamics in terms of the kinetic equation (\ref{secular}). 
 Plugging Eq.\ (\ref{eigenvectors}) into the expression for the transition rates 
 in Eq.\ (\ref{transitionrate}), we find that most of the transition
 rates between two different states belonging to two different pairs 
 are zero, except for  
\begin{eqnarray}
{\cal L}_{nn,n+1n+1}&=&
{\cal L}_{nn,N-n-1 N-n-1}=
{\cal L}_{N-n N-n,n+1n+1}\nonumber\\&=&
{\cal L}_{N-n N-n,N-n-1 N-n-1}
=\frac{\gamma_s}{4} (n+1)\, ,\nonumber\\
{\cal L}_{n+1n+1,nn}&=&
{\cal L}_{n+1n+1,N-n N-n}=
{\cal L}_{N-n-1 N-n-1,nn}\nonumber\\&=&
{\cal L}_{N-n-1 N-n-1,N-n N-n}
=\frac{\gamma_s}{4} (N-n)\, ,\nonumber\\
 {\cal L}_{N/2 N/2, N/2\pm1N/2\pm1}
& = &\frac{\gamma_s}{4} (N+2) \qquad {\rm (for} \,\, N \,\, {\rm  even)} \, ,\nonumber\\
 {\cal L}_{N/2\pm1 N/2\pm1, N/2N/2}
& = & \frac{\gamma_s}{4} N\qquad {\rm (for} \,\, N \,\, {\rm  even).} 
\label{multiphononrates}
\end{eqnarray}
The transition rates between states belonging to the same pair are
zero with the exception  ${\cal L}_{(N-1)/2 (N-1)/2, (N+1)/2 (N+1)/2}
=\gamma_s (N+1)/8$. 

\begin{figure}[h]
\begin{center}
\vspace*{25mm}
\epsfig{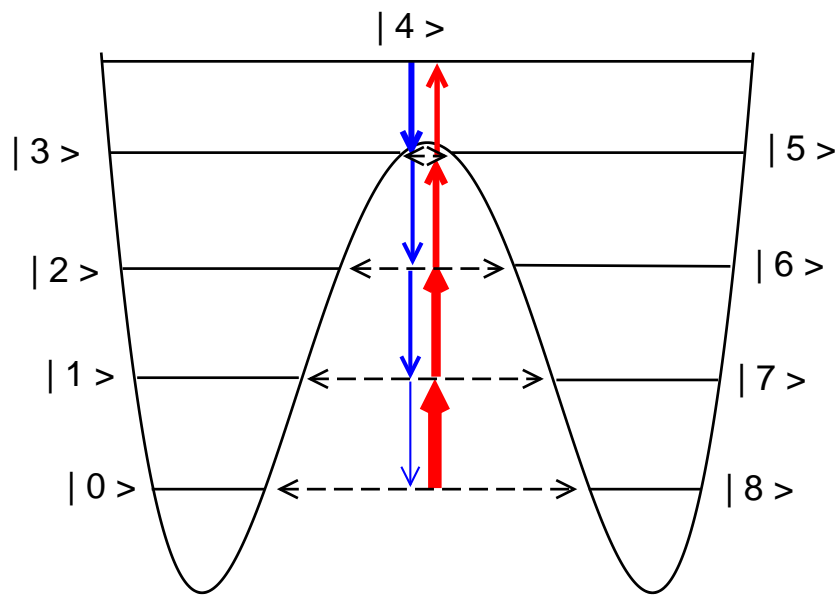}
\caption{Schematic view of the quasipotential and localized  states in 
the rotating frame for the case $N=8$. Shown are the pairs of harmonic oscillator states
consisting of $|n\rangle$ and  $|N-n\rangle$ each of which is
localized in one of the two wells. The corresponding 
quasienergy states $|\phi_{n}\rangle$ and $|\phi_{N-n}\rangle$ 
are a superposition of the two localized states, see text. The 
horizontal arrows indicate the multiphonon transitions between the two
quasienergy states. The vertical arrows mark the bath-induced
transitions with their thickness 
being proportional to the transition rate. \label{fig.0}}
\end{center}
\end{figure}

The dynamics can be illustrated with a simple
analogy to a double-well potential. Each partner of the 
pair $|\phi_n\rangle$ and $|\phi_{N-n}\rangle$ of the quasienergy
states  consists 
of a superposition of two harmonic oscillator states  $|n\rangle$ and 
$|N-n\rangle$ which are the approximate eigenstates of the static
anharmonic potential in the regime of weak nonlinearity. 
In our simple picture, $|n\rangle$ and $|N-n\rangle$ should be
identified with two localized states in the two wells of the quasienergy
potential, see Fig.\ \ref{fig.0} for illustration. 
Note that a quasipotential can be obtained by writing the RWA
Hamiltonian in terms of the two canonically conjugated variables
${\cal X}$ and ${\cal P}$ \cite{Dykman05}. 
The right/left well should be identified with the
internal/external part  of the quasieneryg surface shown in Ref.\  
  \cite{Dykman05}.

In the figure, we
have chosen $N=8$. Within our analogy, the states 
$|0\rangle, |1\rangle, ... , 
|N/2 -1 \rangle$ are localized in one (here, the left) well,  
while $|N\rangle, |N-1\rangle, ... , 
|N/2 + 1 \rangle$  are localized in the other well (here, the right). 
The fact that the true quasienergy states are superpositions of the
two localized states is illustrated by a horizontal arrow representing
tunneling. 

 From Eq.\  (\ref{multiphononrates}) follows that a bath-induced
 transition is only possible between states belonging to two different
 neighboring pairs. As discussed after Eq.\ (\ref{transitionrate}),
 the only contribution to the transition rates come from nearby
HO states. In our case, we consider only spontaneous emission which 
 corresponds to 
 intrawell transitions induced by the bath. This is shown
 schematically in Fig.\ \ref{fig.0} by the vertical arrows with their 
 thickness being proportional to the transition rates.  
 We emphasize that 
  the bath-induced transitions occur towards lower
 lying HO states. Consequently, in our picture, spontaneous decay
 happens in the left well downwards but in the right well upwards.
 
 The driving field excites the transition
 from $|0\rangle$ to $|N\rangle$ while the bath  generates
 transitions between HO states towards lower energies according to  
 $|N\rangle \rightarrow |N-1\rangle\rightarrow ... \rightarrow
 |0\rangle$ when only spontaneous emission is considered.  
 
 As a consequence, the ratio of  
 the occupation numbers of two states  belonging to two neighboring pairs is  
 simply given by the ratio of the corresponding transition rates
 according to 
\begin{eqnarray}
\varrho_{nn}&=&\varrho_{N-n N-n} \, ,\nonumber\\
\frac{\varrho_{nn}}{\varrho_{n+1n+1}}&=&
\frac{{\cal L}_{nn,n+1n+1}}{{\cal L}_{n+1n+1,nn}}=\frac{n+1}{N-n}\,.
\label{secsolution}
 \end{eqnarray}
 %%%%%%%%%\end{figure}
Hence,  
  the unpaired  state $|\phi_{N/2}\rangle$ (for $N$ even) or the states  
  $|\phi_{(N-1)/2}\rangle$ and $|\phi_{(N+1)/2}\rangle$ 
  (for N odd)  are the states with the largest 
  occupation probability. By iteration, one finds 
\begin{eqnarray}
\varrho_{N/2}& = &\left(1+2\sum_{n=1}^{N/2}\prod_{k=0}^{n-1}\frac{N-2k}{N+2+2k}\right)^{-1}
= 0.5,0.37,0.31,0.27,\dots\quad \nonumber \\
& & \mbox{\hspace{37ex}} {\rm for}\quad N=2,4,6,8,\dots \, ,
\end{eqnarray}
and 
\begin{eqnarray}
\varrho_{(N\mp1)/2}&=&\left(2+2\sum_{n=1}^{(N-1)/2}\prod_{k=0}^{n-1}
\frac{N-1-2k}{N+3+2k}\right)^{-1}= 0.37,0.31,0.27,0.25,\dots\quad \nonumber \\
& & \mbox{\hspace{37ex}} {\rm for} \quad N=3,5,7,9,\dots 
\end{eqnarray}

\subsubsection{Density matrix around the resonance \label{subsub}}

So far, we have discussed the dynamics exactly at resonance. Next, we
consider the situation around the resonance and for an 
increased coupling to the bath. Therefore, we  compute the 
stationary solution using the conditions in Eq.\ (\ref{psecular}) 
and the general leading order solution for the quasienergy states
given in Eq.\ (\ref{eigenvectors}). The expressions for the rates 
which are modified compared to before follow
straightforwardly and are given in the Appendix. 
Similarly, the only three equations which change compared to the
previous situation are also presented there. 
These equations can be straightforwardly solved by
\begin{eqnarray}
\fl\varrho_{00}&=&\left[\frac{1}{N}\cot^2\frac{\theta}{2}+\frac{N}{2}
\left(\frac{\gamma_s}{\Omega(\Delta)}\right)^2
\cos^2\frac{\theta}{2}\left(1+\frac{1}{2}\tan^2\frac{\theta}{2}
+\frac{1}{2}\cot^2\frac{\theta}{2}\right)\right]\varrho_{11} \, ,
\nonumber \\
\fl\varrho_{NN}&=&\left[\frac{1}{N}\tan^2\frac{\theta}{2}
+\frac{N}{2}\left(\frac{\gamma_s}{\Omega(\Delta)}\right)^2
\sin^2\frac{\theta}{2}\left(1+\frac{1}{2}\tan^2\frac{\theta}{2}
+\frac{1}{2}\cot^2\frac{\theta}{2}\right)\right]\varrho_{11} \, ,
\nonumber \\
\fl\varrho_{N0}&=&\sin\frac{\theta}{2}\cos\frac{\theta}{2}\left(\frac{N}{2}
\frac{\gamma_s}{\Omega(\Delta)}-i\right)\frac{\gamma_s}{\Omega(\Delta)}
\left(1+\frac{1}{2}\tan^2\frac{\theta}{2}
+\frac{1}{2}\cot^2\frac{\theta}{2}\right)\varrho_{11}\,.
\end{eqnarray}
%\begin{figure}[h]
%%%%%%\end{figure}
Away from  the resonance ($|\theta|\ll 1$), the density matrix follows
as 
\begin{equation}
\varrho\simeq|\phi_0\rangle\langle\phi_0|\simeq|0\rangle\langle 0|\,.
\label{away}
\end{equation}
In the limit of strong coupling  ($\gamma_s\gg\Omega(\Delta)$), one
finds  
\begin{equation}
\fl\varrho\simeq \cos^2\frac{\theta}{2}|\phi_0\rangle\langle\phi_0|+\sin^2
\frac{\theta}{2}|\phi_N\rangle\langle\phi_N|
+\sin\frac{\theta}{2}\cos\frac{\theta}{2}(|\phi_0\rangle\langle\phi_N|
+|\phi_N\rangle\langle\phi_0|)=|0\rangle\langle0|\, , 
\label{strong}
\end{equation}
for any $\theta$. This nicely illustrates that when the coupling to
the bath is strong enough, the possibility of resonant tunneling between 
$|0\rangle$ and $|N\rangle$  is destroyed and a trivial asymptotic
state results. This is true even if  tunneling
transitions between the other states are possible. Moreover, this also
shows that moving away from resonance also suppresses multiphonon
tunneling transitions.  In other words, the only requirement for the
multiphonon transition to occur in the stationary limit 
is the possibility of the tunneling 
transition $|0\rangle \rightarrow |N\rangle$.  

\subsubsection{Lineshape around the resonance}

\begin{figure}[h]
\begin{center}
\vspace*{25mm}
\epsfig{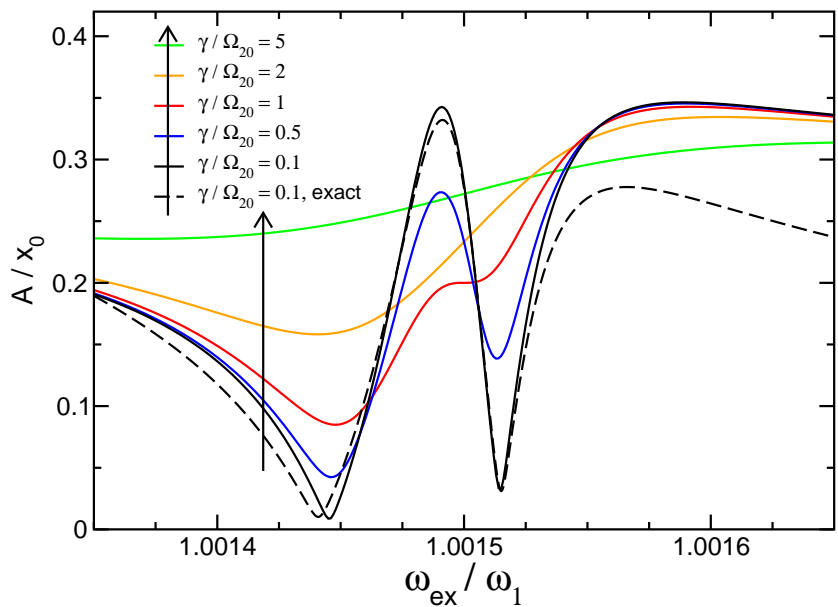}
\caption{Nonlinear response at the two-phonon resonance $N=2$, 
where $\omega_{\rm ex} = \omega_1 + 3\nu/2$ for the
parameters $f=10^{-4}$ and $\nu=10^{-3}$ (in dimensionless units) for
different values of the damping.
 \label{fig.2}}
\end{center}
\end{figure}

Within our partial secular approximation, the lineshape of the 
oscillator nonlinear response  given in Eq.\ (\ref{nonlinearres}) 
 reduces to
 \begin{equation}
A= \sqrt{2}x_0
|\sum_{\alpha\beta}\varrho_{\alpha\alpha}A_{\alpha\alpha}
+\varrho_{0N}A_{0N}+\varrho_{N0}A_{N0}| \, .
\label{responsepsecular} 
 \end{equation}
The leading order is given by the zero-th order expression for 
$\varrho$ and the first-order expressions for $A_{\alpha\alpha}, A_{N0}$ and 
$A_{0N}$. In order to compute these matrix elements,  
we determine the first order eigenvectors using Van Vleck perturbation
theory according to 
\begin{equation}
|\phi_0\rangle_1=e^{i\varepsilon S_1}|\phi_0\rangle_0 \, ,
\end{equation}
where $S_1$ is the first order component in the expansion of 
$S$ with respect to $\varepsilon$ given in Eq.\ (\ref{VanVlecktrans}). 
The matrix elements of its off-diagonal blocks  are given by 
\begin{eqnarray}
\langle \alpha|S_1|\beta\rangle=-i\frac{\langle \alpha|V|\beta\rangle}
{E_\beta-E_\alpha} \, . 
\end{eqnarray}
Here, $E_\alpha$ are the eigenenergies of the unperturbed Hamiltonian
$H_0$ given in Eq.\  (\ref{ham0}). 
This yields  for $N=2$
\begin{eqnarray}
A_{00}=3\varepsilon\left(1-2\sqrt{2}\sin\frac{\theta}{2}\cos\frac{\theta}{2}\right)\,
,
\quad
A_{22}=3\varepsilon\left(1+2\sqrt{2}\sin\frac{\theta}{2}\cos\frac{\theta}{2}\right)\,
,\nonumber
\\
A_{02}=6\sqrt{2}\varepsilon\cos^2\frac{\theta}{2} \, 
\quad A_{20}=-6\sqrt{2}\varepsilon\sin^2
\frac{\theta}{2}\, , \quad A_{11}=-9\varepsilon \, .
\label{aa}
\end{eqnarray}
The corresponding result for the nonlinear response for $N=2$ is shown
in Fig.\  \ref{fig.2} for the case $\nu=10^{-3}$ and $f=10^{-4}$ for
different values of $\gamma_s/\Omega_{20}$.  For strong damping
$\gamma_s/\Omega_{20}=5$, the resonance is washed out almost
completely. Decreasing damping, a resonant lineshape appears whose
maximum is shifted compared to the resonance condition $\omega_{\rm
ex}=\omega_1+3\nu/2$. Note that the dashed line refers to the result
which includes all orders in $\varepsilon$ and which follows from the
numerical solution of the master equation for an Ohmic bath at
temperature $T=0.1 T_0$.  
The picture which arises for the behavior is the following: 
 For weak damping ($\gamma_s \ll \Omega_{20}$), the equilibrium state is 
a statistical mixture of quasienergy states. At resonance, the most
populated state is $|\phi_1\rangle$ which  
oscillates with a phase difference of $-\pi$ in comparison with the
driving.  This is due to the negative sign of $A_{11}$ in Eq.\ 
(\ref{aa}). Hence, at resonance the overall oscillation of the
observable occurs with a phase difference of $\varphi=-\pi$. 
Far away from resonance, the most
populated state is $|\phi_0\rangle$, see Eq.\ (\ref{away}), which
oscillates in phase with the driving. 
Thus, the overall oscillation occurs in phase, i.e., $\varphi=0$. 
If no off-diagonal element of
the density matrix is populated (which is the case for weak damping), 
the overall phase is either $\varphi=0$ or $\varphi=-\pi$. Hence,
increasing the distance from resonance, the amplitude $A$ has to go
through zero yielding a cusp-like line-shape. 
This implies the existence of a maximum in the response. 
For slightly larger damping, the finite population of the off-diagonal
elements leads to a smearing of the cusp. 
For larger damping, the resonance is washed out completely, as has
been already discussed,  see Eq.\  (\ref{strong}). In this regime, the
oscillation is in phase with the driving. By decreasing the damping,
the population of the out-of-phase state starts to increase near the
resonance resulting in a reduction of the in-phase-phase oscillation
and thus producing a minimum of the response.  

\begin{figure}[h]
\begin{center}
\vspace*{25mm}
\epsfig{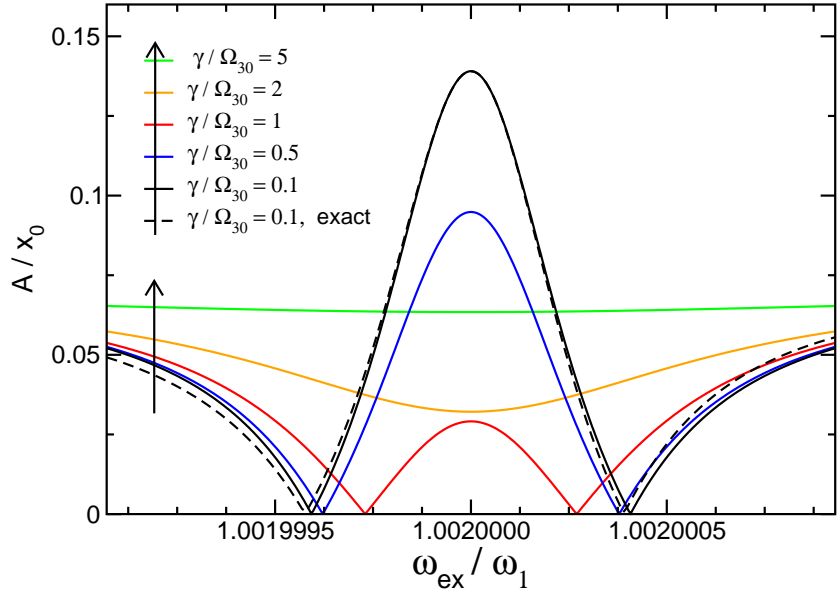}
\caption{Nonlinear response at the three-phonon resonance $N=3$, where
$\omega_{\rm ex} = \omega_1 + 2\nu$ for the
parameters $f=0.5\times 10^{-4}$ and $\nu=10^{-3}$ (in dimensionless units) for
different values of the damping.
 \label{fig.4}}
\end{center}
\end{figure}

This mechanism is effective for a broad range of parameters including 
larger $N$, larger
$\varepsilon$,  and larger temperature $T$,  as also 
shown numerically in Refs.\  \cite{Peano04,Peano05}. 
Note that a calculation up to  first order in $\varepsilon$ for the
density matrix is required for $N$ odd, since the matrix elements
$A_{(N\pm1)/2 (N\pm1)/2}$  have a zero-th order term, in order that the overall result
for $A$ is again of first order in $\varepsilon$. 
Since one obtains more complicated expressions than before, we omit to
present them in their full lengths. In Fig.\  
\ref{fig.4}, we show the behavior for $N=3$ for various
damping constants $\gamma_s/\Omega_{30}$ for the case 
$f=0.5\times 10^{-4}$ and $\nu=10^{-3}$. For a large value for 
$\gamma_s/\Omega_{30}$, the resonance is washed out completely. When
the damping is decreased, a dip appears which corresponds to an
antiresonance. Decreasing the damping further, the antiresonance turns
into a clear resonance. This behavior is opposite to the case $N=1$
 as discussed above, but similar to the case $N=2$.  

\section{Conclusions\label{concl}}

We have studied the nonlinear response of a vibrating nanomechanical beam to a
time-dependent periodic driving. Thereby, a static longitudinal
compression force is included and the system is investigated close to
the Euler buckling instability.  There, the fundamental transverse
mode dominates the dynamics whose amplitude can be described by an effective
single-particle Hamiltonian with a periodically driven anharmonic
potential with a quartic nonlinearity. Damping is modeled
phenomenologically by a bath of harmonic oscillators. We allow for an Ohmic as
well as for a super-Ohmic spectral density and have considered the
regime of weak system-bath coupling. In this regime, the dynamics is
captured by a Born-Markovian master equation formulated in the frame
which rotates with the driving frequency.  The pure driven Hamiltonian
shows avoided level crossings of the quasienergies which correspond to
multiphonon transitions in the resonator. In fact, a transition
between a resonant and an antiresonant behavior at the avoided level
crossings has been found which depends on the coupling to the bath. 
Concentrating to driving
frequencies around the avoided level crossings, the dynamics can be
simplified considerably by restricting  to a few
quasienergy levels. In order to illustrate the basic principles
governing the resonance-antiresonance transition, we investigate the
perturbative regime of weak nonlinearity and weak driving strength.
Then, Van Vleck perturbation theory allows to calculate the quasienergies and the
quasienergy states and an analytic solution of the master equation
becomes possible yielding directly the nonlinear response. For the
one-phonon case, we find a simple expression for the nonlinear
response which displays a Lorentzian resonant behavior for strong
damping. Reducing the damping strength, an antiresonance arises. For
the multiphonon transitions, first an antiresonance 
arises when damping is reduced. For even smaller values of the damping
constant, the antiresonance turns into a resonant peak. This is due to
a subtle interplay of varying populations of quasienergy states which
is affected by the bath.  
 
Finally, a comment on the observability of this effect is in order.
The amplitude $A$ measuring the nonlinear response is of the order of
the oscillator length scale $x_0$. This makes it challenging to
measure the effect directly since the deterministic vibrations are on
the same length scale like the quantum fluctuations. In turn, more
subtle detection strategies have to be worked out, for instance, 
the capacitive coupling of the resonator to a Cooper pair box 
\cite{Armour02,Rabl04} or single electron transistors
\cite{Blencowe00,Mozyrsky04}, the use of squeezed states in this setup
\cite{Rabl04}, the use of a second nanoresonator as a transducer for
the phonon number in the first one \cite{Santamore04}, 
 or the coherent signal amplification by stochastic
resonance \cite{Badzey05}. In any case, the experimental confirmation
of the theoretically predicted effects remains to be provided. 

\section*{Acknowledgments}
This work has been supported by the DFG-SFB/TR 12.

\section*{Appendix: Density matrix around the multiphonon resonance
 \label{app}}
For completeness, we present in this Appendix the calculation of the density
matrix around the multiphonon resonance which is required for Section 
\ref{subsub}. The expressions for the rates
which are modified compared to before are readily calculated to be 
\begin{eqnarray}
{\cal L}_{00,11}&=&{\cal L}_{00,N-1 N-1}=
\frac{\gamma_s}{2} \cos^2\frac{\theta}{2} \, ,\nonumber\\
{\cal L}_{NN,11}&=&{\cal L}_{NN,N-1 N-1}=\frac{\gamma_s}{2}
 \sin^2\frac{\theta}{2} \, ,\nonumber\\
{\cal L}_{11,00}&=&{\cal L}_{N-1 N-1,00}=\frac{\gamma_s}{2}
 N \sin^2\frac{\theta}{2} \, ,\nonumber\\
{\cal L}_{11,NN}&=&{\cal L}_{N-1 N-1,NN}=\frac{\gamma_s}{2}
 N \cos^2\frac{\theta}{2}\nonumber \, ,\\
%{\cal L}_{00,00}&=&-\left(\frac{\omega_{\rm ex}}{\omega}\right)^s N 
%\sin^2\frac{\theta}{2}\nonumber\\
%{\cal L}_{NN,NN}&=&-\left(\frac{\omega_{\rm ex}}{\omega}\right)^s N 
%\cos^2\frac{\theta}{2}\nonumber\\
{\cal L}_{00,0N}&=&{\cal L}_{00,N0}={\cal L}_{NN,0N}={\cal L}_{NN,N0}
={\cal L}_{0N,00}={\cal L}_{N0,00}\nonumber\\
&=&{\cal L}_{0N,NN}={\cal L}_{N0,NN}=
\frac{\gamma_s}{2} N \sin 
\frac{\theta}{2}\cos \frac{\theta}{2} \, ,\nonumber\\
{\cal L}_{11,0N}&=&{\cal L}_{11,N0}={\cal L}_{N-1N-1,0N}
={\cal L}_{N-1N-1,N0}\nonumber\\&=&-
\frac{\gamma_s}{2}
N \sin \frac{\theta}{2}\cos \frac{\theta}{2} \, ,\nonumber\\
{\cal L}_{0N,11}&=&{\cal L}_{N0,11}={\cal L}_{0N,N-1N-1}
={\cal L}_{N0,N-1N-1}\nonumber\\&=&
\frac{\gamma_s}{2} \sin \frac{\theta}{2}\cos \frac{\theta}{2}\,. 
\end{eqnarray}
Similarly, there are only three equations which change compared to the
previous situation. They read  
 \begin{eqnarray}
 0&=&-N\sin^2\frac{\theta}{2}\varrho_{00}+\cos^2\frac{\theta}{2}
 \varrho_{11}+N\cos\frac{\theta}{2}\sin\frac{\theta}{2}
 \varrho_{N0} \, , \nonumber \\
 0&=&-N\cos^2\frac{\theta}{2}\varrho_{NN}+\sin^2\frac{\theta}{2}\varrho_{11}+N\cos\frac{\theta}{2}\sin\frac{\theta}{2}
 \varrho_{N0} \, , \nonumber\\
 0&=&-i\Omega(\Delta)\varrho_{N0}+\frac{\gamma_s}{2}
 \left[-N\varrho_{N0}+\cos\frac{\theta}{2}\sin\frac{\theta}{2}
 (N\varrho_{00}+N\varrho_{NN}+2\varrho_{11})
 \right],
 \end{eqnarray}
with  the quasienergy level splitting   
\[
\Omega(\Delta)=\varepsilon_N-\varepsilon_0=-{\rm sgn}(\Delta)
\left[\left(\frac{\nu(N+1)}{2}N\Delta\right)^2+\Omega^2_{N0}\right]^{1/2}\,.
\] 

\vspace*{25mm}

\end{document}